\documentclass[aps,prl,twocolumn,showpacs,superscriptaddress]{revtex4}
\usepackage{graphicx}
\usepackage{dcolumn}
\usepackage{bm}
\usepackage{amssymb}

\begin{document}

\title{Metamorphosis of a Quantum Hall
Bilayer State into a Composite Fermion Metal}
\author{Biswajit Karmakar}
\affiliation{NEST CNR-INFM and Scuola Normale Superiore, Piazza dei
Cavalieri 7, I-56126 Pisa (Italy)}
\author{Stefano Luin}
\affiliation{NEST CNR-INFM and Scuola Normale Superiore, Piazza
dei Cavalieri 7, I-56126 Pisa (Italy)}
\author{Vittorio Pellegrini}
\affiliation{NEST CNR-INFM and Scuola Normale Superiore, Piazza dei
Cavalieri 7, I-56126 Pisa (Italy)}
\author{Aron Pinczuk}
\affiliation{Dept. of Appl. Phys. and Appl. Math, Dept. of
Physics, Columbia University, New York, New York 10027}
\affiliation{Bell Laboratories, Alcatel-Lucent, Murray Hill, New
Jersey 07974}
\author{Brian S. Dennis}
\author{Loren N. Pfeiffer}
\author{Ken W. West}
\affiliation{Bell Laboratories, Alcatel-Lucent, Murray Hill, New
Jersey 07974}

\date{\today}

\begin{abstract}

Composite fermion metal states emerge in quantum Hall bilayers at
total Landau level filling factor $\nu _T$=1 when the tunneling
gap collapses by application of in-plane components of the
external magnetic field. Evidence of this transformation is found
in the continua of spin excitations observed by inelastic light
scattering below the spin-wave mode at the Zeeman energy. The
low-lying spin modes are interpreted as quasiparticle excitations
with simultaneous changes in spin orientation and composite
fermion Landau level index.
\end{abstract}


\maketitle

Distinct quantum phases of electrons exist in coupled bilayers of double quantum wells
at total Landau  level filling factor $\nu _T = 1$. They result from the
interplay between interlayer and intralayer interactions dictated
by $d/l_B$ ($d$ is the distance between the two layers and
$l_B$ is the magnetic length), and by $\Delta_{SAS}/E_c$, where
$\Delta_{SAS}$ is the tunnelling gap and $E_c = e^2/\varepsilon
l_B$ ($\varepsilon$ is the dielectric constant) is the intra-layer Coulomb
energy \cite{macbook,mur94}.
\par
In the strong interlayer interaction regime, at sufficiently low
$d/l_B$ or large $\Delta_{SAS}/E_c$, the ground states at $\nu _T
=1$ are incompressible quantum Hall (QH) fluids \cite{mur94} (see
Fig.1(a)). In the zero-tunneling limit, these incompressible
states can be described as condensates of interlayer excitons
displaying superfluid-like behavior \cite{kellogg04,simon03}. The
appearance of such highly-correlated QH states occurs when
the Landau levels originating from the left/right
quantum-well subbands are degenerate.
\par
Finite values of $\Delta _{SAS}$ split
the quantum-well states into their symmetric and antisymmetric
combinations. Mean-field approaches have described the QH physics
in these regimes assuming full occupation of the lowest symmetric
Landau level \cite{mac90}. However, recent inelastic light
scattering experiments demonstrated the breakdown of this
mean-field picture by showing that even at $\Delta _{SAS}>0$
inter-layer correlations favor spontaneous occupation of the
excited antisymmetric level \cite{luin05}. These correlated
incompressible states can be described in terms of electron-hole
excitonic pairs across $\Delta _{SAS}$ by making a particle-hole
transformation in the lowest Landau level. These observations
uncover a breakdown of mean-field descriptions that have linked
the stability of these states to the collapse of the magneto-roton energies of
the tunneling charge-density mode \cite{mac90,luin03}.
\par
As the interlayer interaction is reduced either by decreasing
$\Delta_{SAS}/E_c$ or increasing $d/l_B$, the incompressible
excitonic phase is replaced by compressible states as shown in
Fig.1(a). The nature of these compressible states at $\nu _T = 1$
is a subject of current investigation. Several theoretical works
have predicted that composite-fermion (CF) metals occur in these
compressible phases  \cite{simon03,sheng03,macbook}. In the limit
of vanishingly  small inter-layer interactions ($d/l_B\rightarrow
\infty$), the compressible CF metals arise because the two layers,
each at $\nu =1/2$, are decoupled.
\par
CFs were originally introduced to explain the hierarchy of
incompressible states observed in the fractional QH regime
\cite{Jain89}. Due to Chern-Simons gauge fields, CF's with two
attached vortices experience effective magnetic fields $B^*$ =
$B\cdot (1-2\nu)$. As $\nu \to 1/2$ the CF level spacing vanishes
and the CF's are expected to form a new type of metal with a
well-defined Fermi surface.
\par
The interplay between incompressible excitonic and compressible CF
metal phases in bilayers is a topic of great current interest
\cite{stern02, simon03,sheng03,bonesteel96,schl01,
Kim01,Veillette02,Demler01}. On the other hand, experimental
signatures revealing the impact of CF physics in bilayers at $\nu_T
=1$ have been difficult to identify. Recent combined
magneto-transport and nuclear magnetic resonance studies indicate a
tendency to lose spin polarization at the
incompressible-to-compressible phase transition, which suggests
possible links with CF liquids \cite{eisen05,kumada05}.
\par
In this letter we report inelastic light scattering spectra in the
$\nu_T =1$ electron bilayers at finite $\Delta _{SAS}$  that
reveal the metamorphosis of the excitonic incompressible state
into a compressible CF metal.   The tunability of $\Delta _{SAS}$
with changes of the in-plane component of the magnetic field
\cite{hu92} is used here to go across the
incompressible-compressible phase boundary. When the bilayers are
in the excitonic incompressible phase we observe a sharp SW at
$E_z$ due to Larmor theorem. Below a critical value of $\Delta
_{SAS}/E_c$ instead a striking continuum of excitations is seen at
energies smaller than $E_z$.
\par
The continuum of spin transitions seen below $E_z$ is regarded as
evidence of the filling of CF states up to the CF Fermi energy as
shown in Fig.1(b). These spin modes are interpreted as spin-flip
excitations ($SF_{CF}$) of the CF metal in which orientation of
spin and CF Landau level number change simultaneously
\cite{mandal} as displayed in the bottom-right panel of Fig.3.
This low-energy asymmetry of the SW resemble that obtained in CF
metals occurring in single layers at $\nu=1/2$ \cite{duv03}.
Similar $SF_{CF}$ modes are observed at $\nu _T < 1$ and disappear as the
total filling factor approaches the QH state at $\nu_T = 2/3$.
\par
These results demonstrate the emergence of a CF metal phase in
compressible electron bilayers at $\nu _T = 1$. Analysis of the
energy onset of the $SF_{CF}$ continuum (the spin-gap
$E^{CF}_{gap}$ of the bilayer CF metal) raises the possibility
that full spin polarization of the CF metal is lost at a
rather-low critical value of the ratio $\xi = E_{z}/E_c$. $\xi $
measures of the impact of spin in Coulomb interactions linked to
formation of CF quasiparticles and highlights the role played by
the spin degree of freedom in quantum phases of bilayers close to
$\nu_T=1$.
\par
Such studies of low-lying spin excitations by light scattering
methods offer venues to study the multiplicity of states that can
occur near the compressible-incompressible phase boundary at
$\nu_T$ = 1
\cite{simon03,sheng03,stern02,bonesteel96,schl01,Kim01,Veillette02,Demler01}.
The observations of CF ground states at low values of
$\Delta_{SAS}/E_c$ and of excitonic states at larger
$\Delta_{SAS}/E_c$ suggests a subtle competition between these
correlated electronic phases near the phase transition.
\cite{luin06}.
\par
\begin{figure}
\includegraphics%
[width=6.7cm]%
{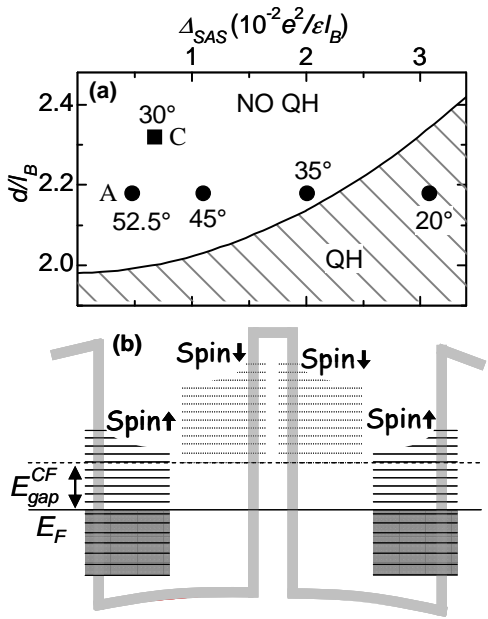}%
\caption{(a) Phase diagram between compressible (NO QH)
and incompressible (QH) phases as
obtained by magneto-transport experiments \cite{mur94}.
Positions of samples studied in this work at different tilt angles are also shown. (b)
Schematic diagram of weakly coupled composite-fermion (CF) metals in the bilayer
system at $\nu_T = 1$. $E_F$ and $E^{CF}_{gap}$ are the Fermi energy and
spin-flip energy gap of the CF metal, respectively.}
\end{figure}
\par
Experiments were carried out in two nominally symmetric
modulation-doped Al$_{0.1}$Ga$_{0.9}$As/GaAs double quantum wells
(DQWs) grown by molecular beam epitaxy and having total electron
densities of $n \sim\ 1.1~-~1.2\times10^{11} ~\textnormal{cm}^{-2}$,
mobilities above $10^6 ~\textnormal{cm}^2 /\textnormal{Vs}$ and
$\Delta _{SAS}$ of 0.36 meV (sample A) and 0.1 meV (sample C). Resonant
inelastic light scattering was performed on samples in a dilution
refrigerator with a base temperature of $\sim\!50$ mK. A tilted
angle geometry with a tilt angle $\theta $ between the total applied magnetic
field $(B_T)$ and the direction perpendicular to the sample surface was exploited. The
non-zero tilt angle gives access to spin excitations when the incident and
scattered photons have crossed polarizations. It also leads
to non-zero components of the in-plane magnetic field that allows to tune
$\Delta _{SAS}$ \cite{luin05, hu92}. Measurements were
performed with laser power densities of
$10^{-3} - 10^{-4} ~\textnormal{W/cm}^{2}$ with a titanium-sapphire laser
tuned close to the DQW optical gap.
\par
Figure 1(a) shows the calculated position of the two samples at the
investigated tilt angles in the $\nu_T$ = 1 phase diagram of
compressible (NO QH) and incompressible (QH) phases. At sufficiently
low values of the inter-layer coupling ($d/l_B \to \infty$,
$\Delta_{SAS}/E_c \to 0$) the ground state can be understood as
composed of CF metals in each of the two layers at $\nu =1/2$, as
shown in Fig.1(b). This state is characterized by a CF Fermi level
$E_F$ and a CF spin-gap $E^{CF}_{gap}$ that separates the highest
occupied CF level with spin-up from the lowest unoccupied CF level
with spin-down. Figure 1(b) refers to the case $E^{CF}_{gap} > 0$
that corresponds to a fully spin polarized CF metal.
\par
\begin{figure}
\includegraphics%
[width=8.5cm]{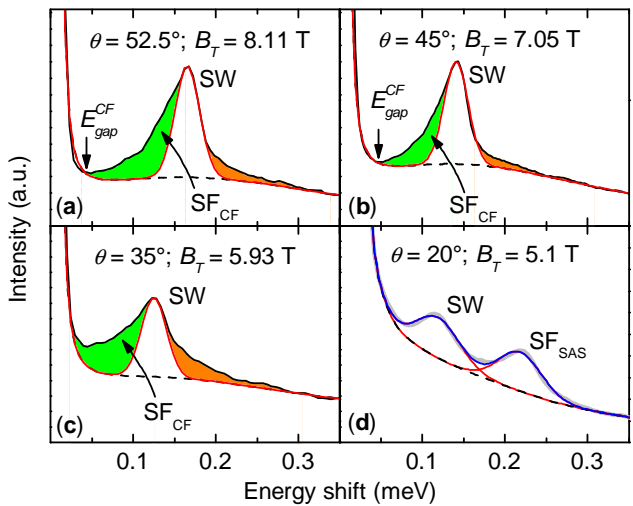}%
\caption{(${a-c}$) Resonant inelastic light scattering spectra of long-wavelength
spin wave (SW) (gaussian line in red) and composite-fermion spin-flip $SF_{CF}$ modes (green area)
at $\nu _T$ = 1 and T = 60mK for different values of tilt angle $\theta$. Dashed line
represents the background due to laser tail and
magneto-luminescence. High-energy scattering (orange area) due to disorder-activated SWs at finite wave-vectors is also shown. For $\theta = 52.5^o$ (a) and $45^o$ (b) the $SF_{CF}$ continuum
(green areas) stops at $E^{CF}_{gap}$ that represents the spin-gap
for CF spin excitations across the Fermi energy (see Fig.1). At
$\theta =35^o$ the continuum extends at energies below the laser
stray light. (d) SW and intersubband spin-flip ($SF_{SAS}$) modes in
the quantum Hall phase at $\theta = 20^{o}$. Red lines are gaussian fits to the
SW and $SF_{SAS}$. Blue line is the fit to the spectrum including the background signal (dashed line). Lorentzian fits of the SW peaks (not shown) lead to similar evaluations of  the $SF_{CF}$ continuum and $E^{CF}_{gap}$.}
\end{figure}
\par
Spin excitation spectra in the $\nu_T = 1$ compressible phase are
displayed in Figs. 2(a-c) for sample A at three different angles
$\theta$. The spectra manifest low-energy  $SF_{CF}$ excitations
(green regions) in addition to the long-wavelength SW peak centered
at $E_z$ due to the Larmor theorem. The SW peak are modeled  by a
gaussian line centered at $E_z$ and with a broadening of $30 \ \mu
eV$. Additional high-energy scattering (in orange) arises because
breakdown of wavevector conservation couples the light to
high-energy SW modes at finite wavevectors. The low-energy
scattering is due to spin excitations because it displays light
scattering selection rules identical to the SW peak. The $SF_{CF}$
feature is assigned to the continuum of CF spin excitations as
suggested by the schematic drawing of CF energy levels with CF
filling factor $\nu_{CF} = \infty$ (where $\nu =
\nu_{CF}/(2\nu_{CF}\pm 1)$), and spin transitions shown in the
right-bottom part of Fig. 3 \cite{duv03}. For $\theta = 45^o$ and $52.5^o$ the
$SF_{CF}$ continuum stops at finite values of $E^{CF}_{gap}$. For
$\theta =35^o$ the continuum extends below the lowest accessible
energy of 30 $\mu eV$ given by the laser stray light which we take
as an indication that the CF metal develops a tendency to lose its
spin polarization at this angle. In order to extract quantitative
information on $SF_{CF}$ continuum and $E^{CF}_{gap}$ the
background signal due to the laser tail and magneto-luminescence
was carefully evaluated. The background (dashed line) is shown in
Figs. 2(a-c) and Fig. 3 for sample A and C, respectively. The same
procedure was adopted for the spectra in the incompressible phase
(Fig. 2(d)).
\par
Below the critical angle $\theta = 35^o$, the $SF_{CF}$ continuum
disappears and a well-defined intersubband spin-flip $SF_{SAS}$
mode across the tunneling gap is observed as shown in Fig.2(d).
Mean-field approaches predict that the SW-$SF_{SAS}$
energy splitting should be equal to $\Delta _{SAS}$. Our previous work
\cite{luin05}, on the contrary, has shown that the splitting is instead less than 
$\Delta _{SAS}$ and determined by the density of excitonic 
electron-hole pairs that exist across $\Delta_{SAS}$.  These excitonic states
are induced by the inter-layer correlations in the
incompressible QH phase \cite{polini}. The phase transition is
seen at the collapse of the SW-$SF_{SAS}$ splitting when the
excitonic density reaches half of the electron density
\cite{luin05}. The results reported here show that the transition
is associated with the appearence of low lying $SF_{CF}$
excitations which demonstrates that close to the critical angle
there is a crossover from an excitonic QH state to a CF metal.
Remarkably enough, the inelastic light scattering measurements
reveal that even very close to the phase boundary, the
compressible state shows characteristic manifestations of CF
quasiparticles despite the residual impact of inter-layer
interactions.
\par
\begin{figure}
\includegraphics%
[width=6cm]%
{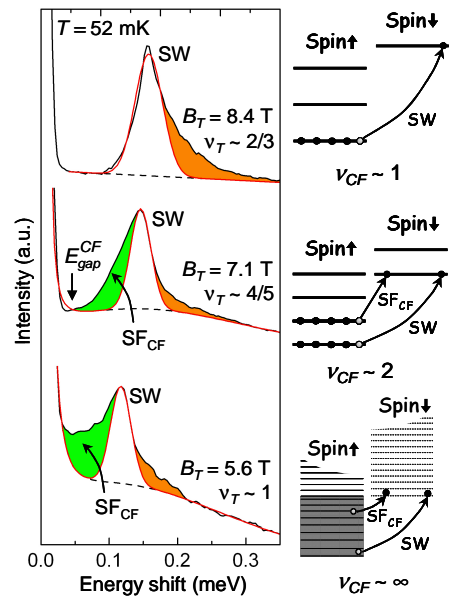}%
\caption{(Left) Spin excitation spectra at three different
values of $\nu _T$ for sample C at a tilt angle of $\theta = 30^o$.
The notations are the same used in Fig.2. The green area that represents the spin-reversal $SF_{CF}$ continuum disappears at $\nu _T$ = 2/3 when one spin-up CF Landau level is expected to be occupied.
(Right) Representation of composite-fermion (CF)
energy levels. Expected occupation of CF quasiparticles and CF filling factor $\nu_{CF}$ are presented for $\nu _T $ = 2/3 and 4/5 and 1. SW and the low-lying $SF_{CF}$ modes are also shown.}
\end{figure}
\par
Figure 3 reports representative spectra of spin excitations at
three different values of $\nu _T$ =1, 4/5 and 2/3 in sample C.
The right part of Fig.3 shows the corresponding CF energy level
schemes, occupations and CF filling fraction $\nu_{CF}$. Similar
to what is reported in Figs 2(a-c) for sample A, the spectral line
shape  at $\nu_T = 1$ shown in Fig.3 is largely affected by
$SF_{CF}$ modes that are signatures of the double-layer CF metal.
The low-lying continuum of $SF_{CF}$ excitations at $\nu_T = 1$
(green region) extends below the lowest accessible energy signaling
possible loss of spin polarization. At $\nu_T \sim 4/5$, instead,
the $E^{CF}_{gap}$ becomes finite demonstrating that at this
magnetic field the system has full spin polarization.
\par
The assignment of the low-energy scattering to spin excitations of
the CF metal is confirmed by the observation that it disappears
when approaching the QH state at $\nu _T = 2/3$. In this regime
the bilayer is composed of two weakly-coupled incompressible $\nu
= 1/3$ QH states in which only the lowest spin-up CF level is
occupied (upper-right part of Fig. 3). No low-lying $SF_{CF}$
excitations can exist in this case consistent with the SW
lineshape shown in Fig.3. The observed SW peak at $\nu_T =2/3$ is
indeed largely asymmetric only at high energy (orange area) as
expected for a weakly-disordered QH state
\par
\begin{figure}
\includegraphics%
[width=6 cm]%
{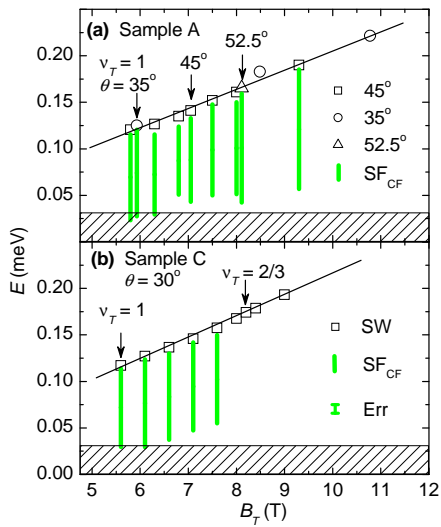}%
\caption{Energies of low-lying spin excitations as a function
of total magnetic field in sample A at different angles (a) and in
sample C (b). Vertical green bars correspond to the contribution of the
composite-fermion $SF_{CF}$ excitations. Open symbols correspond
to the long wavelength limit of the SW. Solid line is a fit with
the relation $E_z = |g_e|\mu _B  B$ yielding $|g_e|=0.41$. Shaded
region at low energy represents the energy range of $\sim 30
~\mu$eV not accessible in our experiments due to the laser tail. 
Typical error bar (Err) in the evaluation of the continuum obtained
by comparing results with Gaussian and Lorentzian lineshapes for the SW is also shown.}
\end{figure}
\par
Figure 4 describes the evolution of the SW peak energy (open dots)
and the extension of the $SF_{CF}$ continuum (vertical bars) in
samples A and C. The angles are chosen such that $\nu _T =1$
corresponds always to the compressible phase (see Fig. 1(a)). In
sample A the $E^{CF}_{gap}$ at $\nu_T  = 1$ remains nearly
constant at higher angles ($> 45^o$) and becomes lower than our
experimental uncertainty between the values of $45^o$ and $35^o$
suggesting a tendency of the CF metal to lose its spin
polarization at low values of $E_z$.
\par
We note that $E^{CF}_{gap}$ is given by the difference between the energy
required to flip a single CF spin and the Fermi energy of the CF sea,
i.e. $E_z + E^{\uparrow \downarrow}-E^{CF}_F = E^{CF}_{gap}$, where
$E^{\uparrow \downarrow}$ arises from the residual interactions
among CFs \cite{duv05}. Both $E^{\uparrow \downarrow}$ and
$E^{CF}_F$ scales with $E_c$ and therefore depends on the
perpendicular component of the magnetic field, while $E_z$ depends
on the total magnetic field. We can thus estimate an upper value of the critical value
of $\xi = E_{z}/E_c$ separating fully spin-polarized from partially
spin-polarized bilayer CF metals. We obtain $\xi _c \le 0.013\pm
0.001$ in sample A. This is lower than the value of $\xi _c
\approx 0.017$ found in single-layer CF metals with similar
densities \cite{duv05}. In a simplified framework, this result can
be understood by invoking a CF effective mass larger than the value
of $\approx 0.4 m_o$ found in single layers ($m_o$ is the free
electron mass) \cite{duv03}. The larger mass may originate from the
impact of inter-layer interactions or reflect the higher
disorder of double layers.
\par
Figure 4(b) highlights that the $SF_{CF}$ continuum in sample C
disappears above $B _T = 8$ T (close to $\nu _T$ = 2/3) as also
shown in Fig.3. Additionally the $E^{CF}_{gap}$ becomes lower than the
experimental uncertainty for magnetic fields below 6T ($E_z \approx
0.13 $meV) suggesting loss of spin polarization in this range of magnetic fields
consistent with data obtained in sample A.
\par
In conclusion, light scattering spectra uncovered a phase
transformation of an excitonic incompressible state into a
compressible (metallic) CF phase as the tunneling gap collapses in
electron bilayers at $\nu_T$ =1. Studies of the CF metal near the
compressible-incompressible phase boundary could offer insights on
interactions involved in the emergence of ground states due the
impact of inter- and intra-layer correlations in electron
bilayers in the quantum Hall regimes.
\begin{acknowledgments}
We are grateful to Steven H. Simon for useful comments.
VP acknowledges the Italian Ministry of Research (FIRB).
AP acknowledges the National Science Foundation under Award Number
DMR-03-52738, the Department of Energy under award
DE-AIO2-04ER46133, and the W. M. Keck Foundation.
\end{acknowledgments}

\end{document}